\documentclass[aps,pre,reprint,superscriptaddress]{revtex4-1}
\usepackage{graphicx}
\usepackage{amsmath}
\usepackage{amssymb}
\usepackage{bm}
\usepackage{physics}
\usepackage{subfigure}
\usepackage[colorlinks=true,linkcolor=blue,urlcolor=blue,citecolor=blue]{hyperref}
\usepackage{times}

\begin{document}
\title{Fidelity Out-of-Time-Order Correlator in the Spin-Boson Model}
\author{Ruofan Chen}
\affiliation{College of Physics and Electronic Engineering, and Center for Computational Sciences, Sichuan Normal University, Chengdu 610068, China}
\date{\today}

\begin{abstract}
  In this article, using the numerically exact time-evolving matrix
  product operators method, we study the fidelity out-of-time-order
  correlator (FOTOC) in the unbiased spin-boson model at zero
  temperature. It is found that after the initial exponential growth
  of FOTOC, the information of the system dynamics will adulterate
  into the FOTOC. This makes the FOTOC an advanced epitome of the
  system dynamics, i.e., the FOTOC shows similar behavior to that of
  system dynamics within a shorter time interval. Eventually the
  progress of the FOTOC is ahead of the system dynamics, which can
  provide a prediction of the system dynamics.
\end{abstract}
\maketitle

\section{Introduction}
The out-of-time-order correlator (OTOC), which was first introduced in
the vertex correction of a current in a superconductor
\cite{larkin1969-quasiclassical} in 1960s, is recently proposed as a
quantum generalization of a classical measure of chaos and to describe
quantum information scrambling in a quantum system
\cite{li2017-measuring,hashimoto2017-out,shen2017-out,syzranov2018-out,garcia-mata2018-chaos,swingle2018-unscrambling,garttner2018-relating,alavirad2019-scrambling,yan2020-recovery,harrow2021-separation,zonnios2022-signatures}.
Recently, it is shown that fidelity OTOC (FOTOC), a specific family of
OTOC, can provide profound insight on quantum scrambling
behavior. This particular type of OTOC has been considered for
studying the multiple quantum coherence spectrum
\cite{garttner2017-measuring, garttner2018-relating}, for quantifying
scrambling in Dicke model \cite{lewis-swan2019-unifying} and quantum
Rabi model \cite{kirkova2022-out}.

The Dicke model \cite{dicke1954-coherence} and quantum Rabi model,
which describe the coupling between spin(s) and a single oscillator
mode, are fundamental models in quantum optics. In condensed matter
physics, there is a fundamental model known as the spin-boson model
\cite{leggett1987-dynamics,weiss1993-quantum}. The spin-boson model is
similar to the Dicke and quantum Rabi models in the sense that it also
describes the coupling between a spin and oscillator modes. The
difference is that in the spin-boson model, the spectrum of oscillator
frequencies is sufficiently dense such that it can be treated
continuous and smooth. The presence of continuous modes means that the
consideration of recurrence phenomena, which is known to be important
in a quantum Rabi like model \cite{eberly1980-periodic}, is excluded.

In this article, using an extension of numerically exact time-evolving
matrix product operators (TEMPO) method
\cite{strathearn2018-efficient,chen2023-heat}, we calculate the FOTOCs
in the unbiased spin-boson model in (sub)ohmic regime at zero
temperature. With continuous oscillator modes, the FOTOCs always
display exponential deviation from unity at the very beginning of the
evolution when the system and bath starting to be entangled. The speed
of the deviation is proportional to the square of the system-bath
coupling strength. That is, with the same bath the FOTOCs coincides
within a very short time despite a factor involving coupling strength
factor. Soon after the information of the system dynamics is
adulterated into the FOTOCs, which makes them deviates from each
other. As the evolution goes on, the FOTOC would contain enough
information of the system dynamics and form an advanced epitome of the
system dynamics.

This article is organized as follows. The definition of the FOTOC and
the model is introduced in Sec. \ref{sec:fotoc}. The method is
presented in Sec. \ref{sec:method}. The FOTOCs beyond and within the
scrambling time are discussed in Sec. \ref{sec:within} and
Sec. \ref{sec:beyond}, respectively. Finally conclusions are given in
Sec. \ref{sec:conclusions}.

\section{Fidelity Out-of-Time-Order Correlator in the Spin-Boson Model}
\label{sec:fotoc}
Here we consider the unbiased spin-boson model
\cite{leggett1987-dynamics,weiss1993-quantum} ($\hbar=1,k_B=1$),
whose Hamiltonian is given by
\begin{equation}
  \begin{split}
    \hat{H}=&\hat{H}_S+\hat{H}_B+\hat{H}_{SB}\\
    =&\frac{\Delta}{2}\hat{\sigma}_x+\sum_k\omega_k\hat{b}^{\dag}_k\hat{b}_k+
    \frac{\hat{\sigma}_z}{2}\sum_kg_k(\hat{b}^{\dag}_k+\hat{b}_k),
  \end{split}
\end{equation}
here $\Delta$ is the tunneling splitting and $\hat{\sigma}_{x/z}$ are
the Pauli matrices. Here $\hat{b}^{\dag}_k$ ($\hat{b}_k$) creates
(annihilates) a boson of state $k$ in bath with frequency $\omega_k$,
which are coupled to the spin via coupling constant $g_k$. The bath is
characterized by a spectral function
\begin{equation}
  J(\omega)=\sum_kg_k^2\delta(\omega-\omega_k)=
  2\alpha\omega^s\omega_c^{1-s}e^{-\omega/\omega_c},
\end{equation}
where $\alpha$ is the coupling strength parameter and $\omega_c$ is
the cutoff frequency of the bath. Here we consider the (sub-)ohmic
case where $0<s\le1$ and a high frequency cutoff $\omega_c=10\Delta$
unless specified otherwise.

The out-of-time-order correlator (OTOC)
\begin{equation}
  \mathcal{F}(t)=\langle\hat{W}^{\dag}(t)\hat{V}^{\dag}\hat{W}(t)\hat{V}\rangle
\end{equation}
can be served as a diagnostic of quantum chaos, where the brackets
denote averaging over the initial state $\ket{\psi_0}$.  Here
$\hat{W}$ and $\hat{V}$ are two initially commuting and Hermitian
operators, and $\hat{W}(t)=e^{i\hat{H}t}\hat{W}e^{-i\hat{H}t}$. The
OTOCs quantify the degree that $\hat{W}(t)$ and $\hat{V}$ fail to
commute at later times due to the time evolution. In fast scramblers,
the quantity $1-\Re\mathcal{F}(t)$ features an exponential growth for
which $1-\Re\mathcal{F}(t)\approx e^{\lambda_Qt}$ before it
saturates. The quantity $\lambda_Q$ is the quantum Lyapunov exponent
associated with quantum chaos.

In this article, we focus on the so-called fidelity OTOC (FOTOC) where
$\hat{V}$ is the projection operator onto the initial state
$\ket{\psi}$, i.e., $\hat{V}=\hat{\rho}(0)=\ketbra{\psi}{\psi}$ and
$\hat{W}=e^{i\xi\hat{A}}$ to be a small perturbation ($\xi\ll1$) for a
Hermitian operator $\hat{A}$. For a pure initial state, we have
$\mathcal{F}(t)=|\bra{\phi}{\hat{W}(t)}\ket{\phi}|^2$. 

The FOTOC $\mathcal{F}(t)$ is a real quantity, therefore we write
$1-\mathcal{F}(t)$ instead of $1-\Re\mathcal{F}(t)$. Expanding
$\mathcal{F}(t)$ in a power series of $\xi$ to the second order yields
\begin{equation}
  1-\mathcal{F}(t)=\xi^2(\langle\hat{A}^2(t)\rangle-\langle\hat{A}(t)\rangle^2)
  =\xi^2\mathrm{var}[\hat{A}(t)],
\end{equation}
where $\mathrm{var}[\hat{A}(t)]$ is the variance of $\hat{A}(t)$. This
relation connects the exponential growth of quantum variances and
quantum chaos. 

In Dicke \cite{lewis-swan2019-unifying} and quantum Rabi
\cite{kirkova2022-out} model, the bath consists of a single mode boson
and the operator $\hat{A}$ is set to be proportional to
$(\hat{b}+\hat{b}^{\dag})$. In our spin-boson model, the bath consists
of a continuous spectrum of bosons, we chose the corresponding
operator to be $\hat{A}=\sum_kg_k(\hat{b}_k+\hat{b}^{\dag}_k)$. The
initial state is $\ket{\psi}=\ket{+,0}$, where the spin is in up state
($\hat{\sigma}_z\ket{+,0}=\ket{+,0}$) and the bath is in its ground
state.

\section{Method}
\label{sec:method}
At the initial time, the total density matrix is in a product form as
$\hat{\rho}(0)=\hat{\rho}_S(0)\hat{\rho}_B(0)$, where
$\hat{\rho}_S(0)$ is the initial system density matrix and
$\hat{\rho}_B(0)$ is the bath density matrix. The bath is in its
ground state, i.e., at zero temperature. With such a product initial
state, the reduced dynamics of the system at a later time
$\hat{\rho}_S(t)=\Tr_B[\hat{\rho}(t)]$ can be formulated as a path
integral via tracing out the bath
\cite{feynman1963-the,weiss1993-quantum,grabert1988-quantum,negele1998-quantum,chen2023-heat}.
Let $\sigma(\tau)$ be the spin path along a Keldysh contour
$\mathcal{C}$
\cite{keldysh1965-diagram,lifshitz1981-physical,kamenev2009-keldysh,wang2013-nonequilibrium}
evolving forward from time $0$ to $t$ then evolving backward to $0$,
then we can write
\begin{equation}
  \label{eq:path-integral-rho}
  \hat{\rho}_S(t)=\int\mathcal{D}[\sigma(\tau)]K[\sigma(\tau)]I[\sigma(\tau)],
\end{equation}
where the integral over $\mathcal{D}[\sigma(\tau)]$ means summation
over all possible paths.  Here $K[\sigma(\tau)]$ is the free system
propagator and $I[\sigma(\tau)]$ is the Feynman-Vernon influence
functional which captures the effects of the bath on the system.

The contour-ordered Green's function of the free bath is defined as
$G_k(\tau',\tau'')=\langle
T_{\mathcal{C}}\hat{b}_k(\tau')\hat{b}^{\dag}_k(\tau'')\rangle$, where
$T_{\mathcal{C}}$ is the contour-ordering operator. Let
$G_{\omega}(\tau',\tau'')$ denote the Green's function
$G_k(\tau',\tau'')$ when $\omega_k=\omega$, then the influence
functional can be written as
\begin{equation}
  I[\sigma(\tau)]=e^{-\int_{\mathcal{C}}\dd{\tau'}\int_{\mathcal{C}}\dd{\tau''}\sigma(\tau')\varGamma(\tau',\tau'')\sigma(\tau'')},
\end{equation}
where
\begin{equation}
  \varGamma(\tau',\tau'')=\int\dd{\omega}J(\omega)G_{\omega}(\tau',\tau'').
\end{equation}

The quantity $\langle\hat{W}(t)\rangle$ can be obtained from a
modified reduced density matrix
\begin{equation}
  \hat{\rho}^{\xi}_S(t)=\Tr_B[\hat{\rho}(t)e^{i\xi\hat{A}}]
\end{equation}
via $\langle\hat{W}(t)\rangle=\Tr_S\hat{\rho}_S^{\xi}$. This
modified reduced density matrix can be represented as a path integral
for which
\begin{equation}
  \label{eq:path-integral-rho-xi}
  \hat{\rho}^{\xi}_S(t)=\int\mathcal{D}[\sigma(\tau)]K[\sigma(\tau)]I[\sigma(\tau)]X_{\xi}[\sigma(\tau)],
\end{equation}
where
\begin{equation}
  X_{\xi}[\sigma(\tau)]=e^{\xi\int_{\mathcal{C}}
    \dd{\tau}[\sigma(\tau)\varGamma(\tau,t^-)+\varGamma(t^+,\tau)\sigma(\tau)]}.
\end{equation}
Here $t^{\pm}$ means time $t$ on the forward (backward) branch of
contour $\mathcal{C}$.

Depending on the parameters on the forward or backward branch, the
contour-ordered Green's function $G_k(\tau',\tau'')$ can be split into
four blocks $G_k^{\pm\pm}(t',t'')$.  By doing this we bring the
influence functional $I[\sigma(\tau)]$ into normal time axis as
$I[\sigma^{\pm}(t)]$. Similarly, the function
$\varGamma(\tau',\tau'')$ can be also split into four blocks
$\varGamma^{\pm\pm}(t',t'')$, and the quantity $X_{\xi}[\sigma(\tau)]$
can be written in normal time axis as $X_{\xi}[\sigma^{\pm}(t)]$.

For numerical evaluation, the influence functional can be discretized
via quasi-adiabatic path-integral (QUAPI) method
\cite{makarov1993-tunneling,makri1995-numerical,dattani2012-analytic}.
Split $t$ into $N$ pieces for which $t=N\delta t$ and the path
$\sigma^{\pm}(t)$ into intervals of equal duration for which
$\sigma^{\pm}(t')=\sigma^{\pm}_j$ for
$(j-\frac{1}{2})\delta t<t'<(j+\frac{1}{2}\delta t)$. This splitting
corresponds to first order Trotter-Suzuki decomposition
\cite{trotter1959-product,suzuki1976-generalized} whose error is about
$O(\delta t^2)$. It is easy to adapt the higher order symmetrized
Trotter-Suzuki decomposition
\cite{makri1995-numerical,makri1995-tensor-i,dattani2012-analytic}
which reduces the error to $O(\delta t^3)$. All the numerical results
in this article use the higher order symmetrized decomposition, but
for ease of exposition we use the form of first order
decomposition. With this splitting, the influence functional
$I[\sigma^{\pm}(t)]$ is discretized as
\begin{equation}
  I[\qty{\sigma^{\pm}_k}]=e^{-\sum_{j=0}^N\sum_{k=0}^j(\sigma_j^+-\sigma_j^-)(\eta_{jk}\sigma_k^+-\bar{\eta}_{jk}\sigma_k^-)},
\end{equation}
where $\eta_{jk}$ is a complex number and $\bar{\eta}_{jk}$ is its
complex conjugate. For $j\ne k$ we have
\begin{equation}
  \eta_{jk}=\int_{(j-\frac{1}{2})\delta t}^{(j+\frac{1}{2})\delta t}\dd{t'}\int_{(k-\frac{1}{2})\delta t}^{(k+\frac{1}{2})\delta t}\dd{t''}C(t'-t''),
\end{equation}
and for $j=k$
\begin{equation}
  \eta_{jj}=\int_{(j-\frac{1}{2})\delta t}^{(j+\frac{1}{2})\delta t}\dd{t'}\int_{(j-\frac{1}{2})\delta t}^{t'}\dd{t''}C(t'-t''),
\end{equation}
where $C(t)$ is the autocorrelation function.

As mentioned in Ref. \cite{chen2023-heat}, to be consistent with QUAPI
method, the variable $\xi$ need to be replaced by a segment as
\begin{equation}
  \xi\to\frac{1}{\delta t}\int_{(N-\frac{1}{2})\delta t}^{(N+\frac{1}{2})\delta t}\xi(t')\dd{t'}
\end{equation}
with $\xi(t')=\xi$. Therefore $X_{\xi}[\sigma^{\pm}(t)]$ should be
discretized as
\begin{equation}
X_{\xi}[\qty{\sigma^{\pm}_k}]=e^{\xi\sum_{j=0}^N(\sigma_j^+\gamma^{+-}_{jN}-\sigma_j^-\gamma^{--}_{jN}+\gamma^{++}_{Nj}\sigma^+_j-\gamma^{+-}_{Nj}\sigma^-_j)},
\end{equation}
where for $j\ne N$
\begin{equation}
  \gamma^{\pm\pm}_{jN}=\frac{1}{\delta t}\int_{(N-\frac{1}{2})\delta t}^{(N+\frac{1}{2})\delta t}\dd{t'}
  \int_{(j-\frac{1}{2})\delta t}^{(j+\frac{1}{2})\delta t}\dd{t''}\varGamma^{\pm\pm}(t'-t''),
\end{equation}
and for $j=N$
\begin{equation}
  \gamma^{\pm\pm}_{NN}=\frac{1}{\delta t}\int_{(N-\frac{1}{2})\delta t}^{(N+\frac{1}{2})\delta t}\dd{t'}
  \int_{(N-\frac{1}{2})\delta t}^{t'}\dd{t''}\varGamma^{\pm\pm}(t'-t'').
\end{equation}

The memory time of bath is finite for which $C(t)$ and
$\varGamma^{\pm\pm}(t)$ decay to zero for sufficiently large $t$.
Therefore the corresponding $\eta_{jk}$ and $\gamma^{\pm\pm}_{jk}$ can
be truncated when $\abs{j-k}$ is larger than a positive integer
$\Delta k_{\mathrm{max}}$. This is the key ingredient of QUAPI method,
which enables us to simulate the long time evolution of
$\hat{\rho}_S(t)$ iteratively in a tensor multiplication manner.

The iterative process can be implemented in the language of matrix
product state (MPS) and matrix product operator (MPO), which gives a
MPS representation of reduced density matrix
\eqref{eq:path-integral-rho}. This yields the so-called time-evolving
matrix product operators (TEMPO) algorithm
\cite{strathearn2018-efficient}. The TEMPO method can employ the
standard MPS compression algorithm \cite{schollwoeck2011-density}
during the iterative process, makes it computationally efficient and
yet numerically exact. The tensor $X_{\xi}[\qty{\sigma^{\pm}_k}]$ can
be easily represented as a MPO, and applying it to the MPS
representation of Eq. \eqref{eq:path-integral-rho} yields the modified
reduced density matrix \eqref{eq:path-integral-rho-xi}.

In this article, we use the singular value decomposition (SVD)
algorithm to compress the MPS. This operation is done by truncating
all singular values $\lambda<\varepsilon\lambda_{\mathrm{max}}$, where
$\lambda_{\mathrm{max}}$ is the largest singular value and
$\varepsilon$ is a convergence parameter. The FOTOC $\mathcal{F}(t)$
is much more numerically sensitive than the polarization
$P(t)=\langle\sigma_z(t)\rangle$, therefore we need to adopt a fine
$\varepsilon=10^{-11}$. The value of perturbation $\xi$ is chosen to
be $\xi=10^{-3}$.

\section{Beyond the Scrambling Time}
\label{sec:beyond}

\begin{figure}[htbp]
\centerline{\includegraphics[]{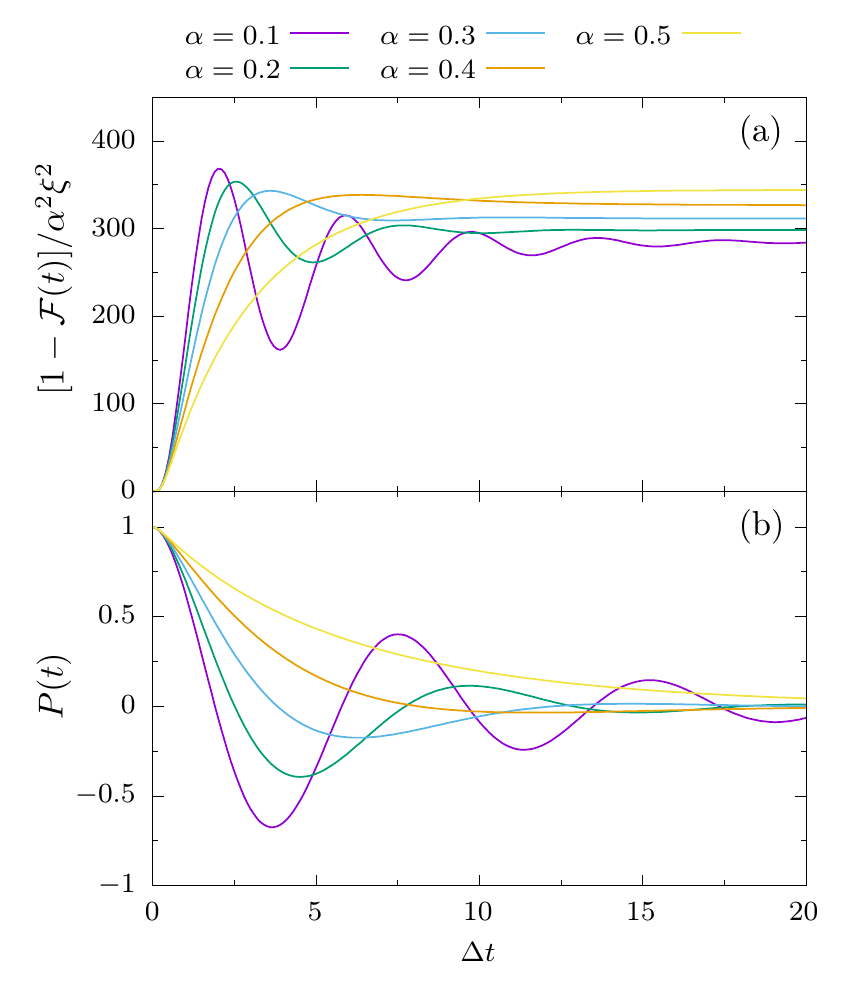}}
\caption{Some typical (a) FOTOCs and (b) polarizations in ohmic
  $s=1.0$ regime with different coupling strengths.}
\label{fig:01}
\end{figure}

\begin{figure}[htbp]
\centerline{\includegraphics[]{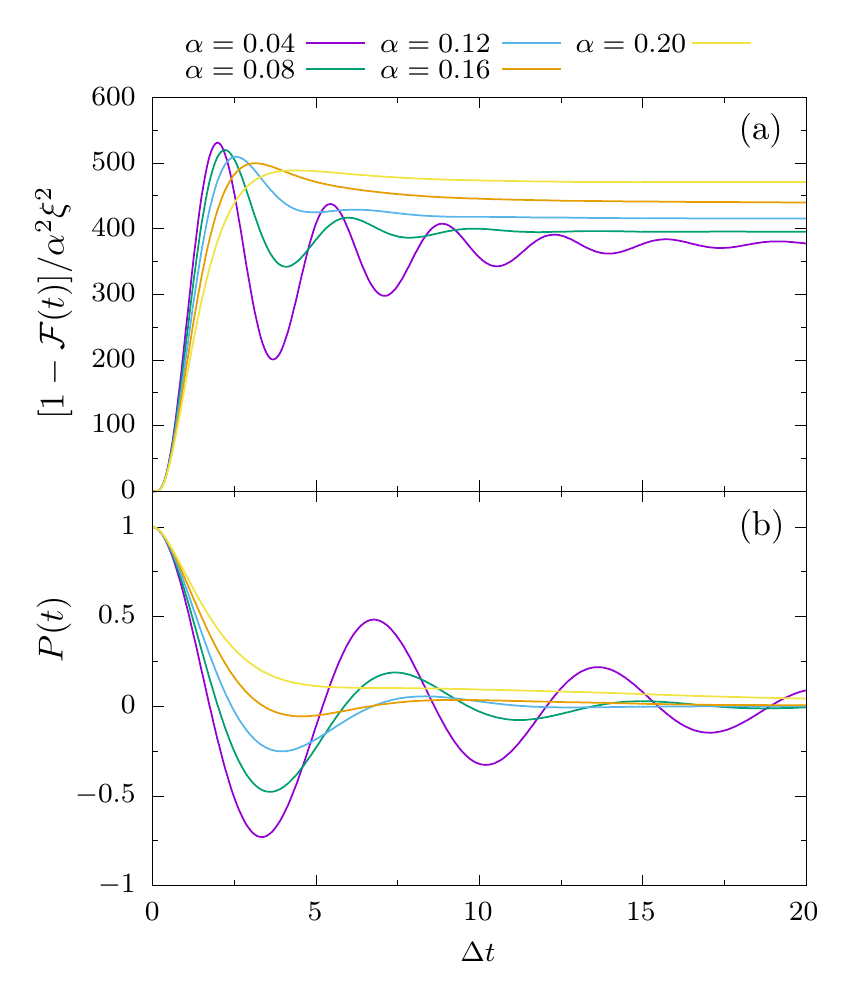}}
\caption{Some typical (a) FOTOCs and (b) polarizations in subohmic
  $s=0.7$ regime with different coupling strengths.}
\label{fig:02}
\end{figure}

Figures \ref{fig:01} and \ref{fig:02} show some typical FOTOCs and
corresponding polarizations $P(t)=\langle\hat{\sigma}_z(t)\rangle$ in
ohmic ($s=1.0$) and subohmic ($s=0.7$) regimes with different coupling
strength $\alpha$. The time step is $\delta t=0.1$ in the ohmic case
and is $\delta t=0.06$ in the subohmic case. The magnitude of
$[1-\mathcal{F}(t)]/\xi^2$ roughly scales with $\alpha^2$, thus we
show the scaled $[1-\mathcal{F}(t)]/\alpha^2\xi^2$ rather than
$[1-\mathcal{F}(t)]/\xi^2$. It can be seen that their early behaviors
look similar for which all these curves start to grow significantly at
the beginning of the evolution and then saturate. After the
saturation, they show different long time behaviors.

Usually the scrambling time is identified as the time when
$[1-\mathcal{F}(t)]/\alpha^2\xi^2$ reaches its first local
maximum. However, there are some situations, e.g., in
Fig. \ref{fig:01}(a) the curve just grows monotonically with
relatively strong coupling $\alpha=0.5$, where no local maximum can be
identified within a short time. In this case, we may roughly identify
the scrambling time as the time when the curve decelerates most. Here
we try not to identify the scrambling time exactly and just use it as
a vague concept to distinguish short and long times. The reason for
doing this will be shown in the next section.

It is well known \cite{leggett1987-dynamics,weiss1993-quantum} that at
zero temperature, in ohmic regime the polarization $P(t)$ shows damped
coherent oscillations with weak coupling and incoherent decay at
stronger dissipation, see Fig. \ref{fig:01}(b). The transition occurs
at $\alpha\approx0.5$. Similar phenomena happen in the subohmic regime
with a smaller transition point $\alpha$, see
Fig. \ref{fig:02}(b). After the scrambling time, the behaviors of
FOTOCs show similarity to the corresponding polarization dynamics in
both ohmic (Fig. \ref{fig:01}) and subohmic (Fig. \ref{fig:02})
regimes. This reminds us that the FOTOCs may contain information of
polarization dynamics.

\begin{figure}[htbp]
\centerline{\includegraphics[]{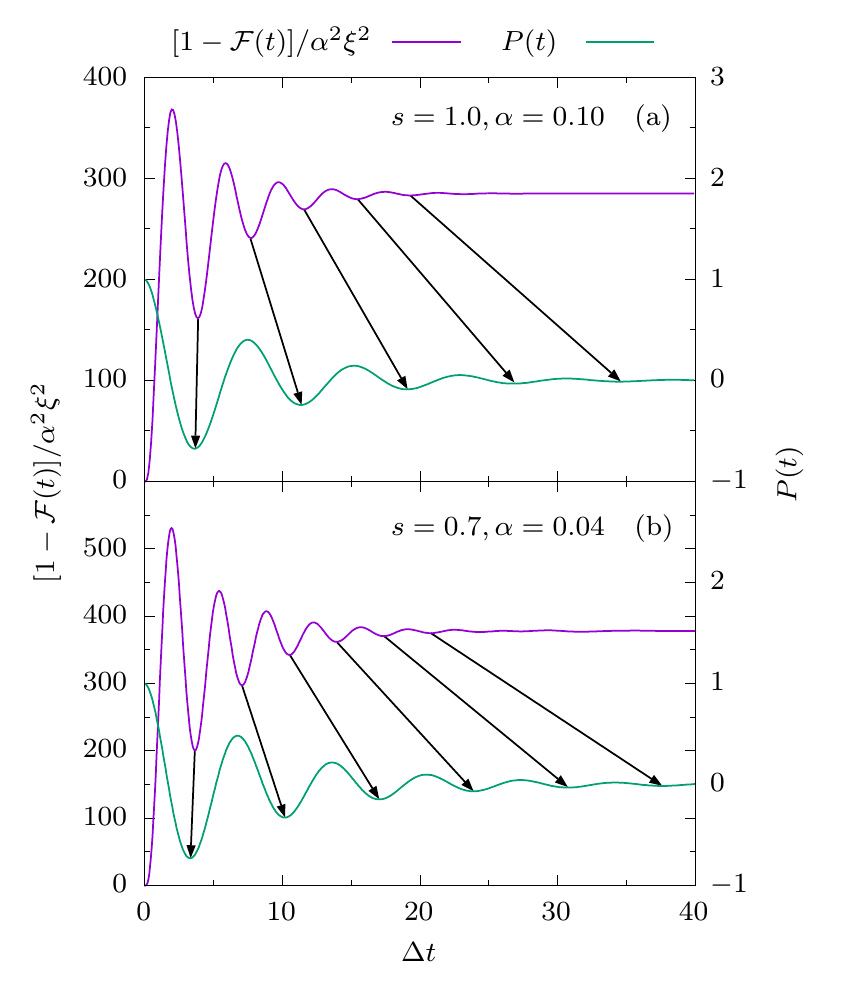}}
\caption{A detailed comparison between the FOTOCs and the
  polarizations in the ohmic regime. The coupling strength is (a)
  $\alpha=0.1$ and (b) $\alpha=0.2$.}
\label{fig:03}
\end{figure}

We choose the ohmic $s=1.0,\alpha=0.1$ [Fig. \ref{fig:03}(a)] and the
subohmic $s=0.7,\alpha=0.2$ [Fig. \ref{fig:03}(b)] cases to
demonstrate the similarity between the FOTOCs and the
polarizations. We compare the locations of local minimums of the FOTOC
and the polarization, and denote the location of $n$th minimum of them
as $t_{\mathcal{F}}^n$ and $t_P^n$ respectively.

In Fig. \ref{fig:03}(a), the polarization $P(t)$ shows damped coherent
oscillation with coupling strength $\alpha=0.1$. Here we compare the
first fifth minimums. The first minimums appear at
$t_{\mathcal{F}}^1=3.9$ and $t_P^1=3.7$ for which the minimum of
polarization is slightly ahead of that of FOTOC. The second minimums
appear at $t_{\mathcal{F}}^2=7.7$ and $t_P^2=11.4$ for which
polarization is now behind the FOTOC. The rest locations of minimums
are
$t_{\mathcal{F}}^3=11.6,t_{\mathcal{F}}^4=15.5,t_{\mathcal{F}}^5=19.3$
and $t_P^3=19.1,t_P^4=26.84,t_P^5=34.56$, from which it can be seen
that the difference between $t_{\mathcal{F}}^n$ and $t_P^n$ becomes
larger along with larger $n$. The time interval between
$t_{\mathcal{F}}^1$ and $t_{\mathcal{F}}^5$ is $15.4$ and that between
$t_P^1$ and $t_P^5$ is $30.86$, which means the FOTOC go through a
similar process with only half the time than polarization.

In Fig. \ref{fig:03}(b), the locations of minimums are
$t_{\mathcal{F}}^1=3.66,t_{\mathcal{F}}^2=7.08,t_{\mathcal{F}}^3=10.56,t_{\mathcal{F}}^4=13.98,t_{\mathcal{F}}^517.4,t_{\mathcal{F}}^6=20.76$
and
$t_P^1=3.36,t_P^2=10.2,t_P^3=17.04,t_P^4=23.88,t_P^5=30.72,t_P^6=37.56$. The
progress of the FOTOC is also behind the polarization at first, then
takes the lead at later time. The interval
$t_{\mathcal{F}}^6-t_{\mathcal{F}}^1=17.1$ is also much shorter than
$t_P^6-t_P^1=34.2$. Therefore we can conclude that the FOTOC forms an
epitome of system dynamics, in the sense that it goes through a
similar process in a shorter time than system dynamics. In addition,
since $t_{\mathcal{F}}^n$ is usually ahead of $t_P^n$ the FOTOC also
gives a prediction of the system dynamics.

For large system bath coupling situation, e.g., $\alpha=0.5$ in
Fig. \ref{fig:01}, the polarization shows incoherent decay and there
is no local minimum. In this case, the first local maximum of the
corresponding FOTOC disappears and the FOTOC shows incoherent
growth. The FOTOC reaches the steady state before the polarization,
therefore we may still say that the FOTOC gives an epitome of the
polarization.

\section{Within the Scrambling Time}
\label{sec:within}

\begin{figure}[htbp]
\centerline{\includegraphics[]{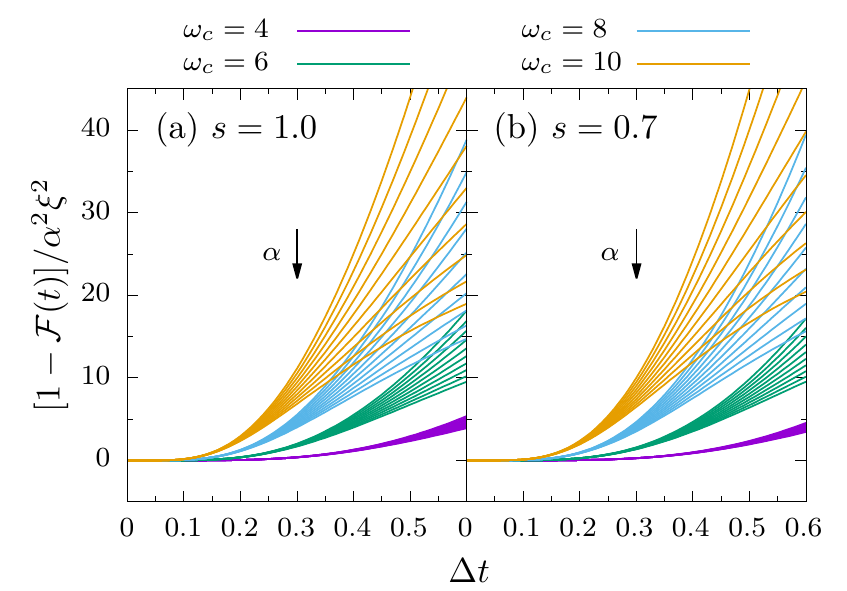}}
\caption{FOTOCs at a short time with different cutoff frequency
  $\omega_c$ and coupling strength $\alpha$ in (a) ohmic ($s=1.0$)
  regime and (b) subohmic ($s=0.7$) regime. The curves with same color
  corresponds to same $\omega_c$ and the coupling strength is varied
  from $\alpha=0.1$ to $\alpha=1.0$ by step $0.1$.}
\label{fig:04}
\end{figure}

Let us take a closer look at the FOTOCs within the scrambling time. In
Fig. \ref{fig:04}, we show the FOTOCs within a very short time with a
fine time step $\delta t=0.01$. In the figure, the curves with the
same color correspond to various coupling strengths but same
$\omega_c$.

It can be seen that with scaling factor $\alpha^{-2}$, the FOTOCs with
same $\omega_c$ almost coincide at the beginning of the evolution, and
then start to deviate soon after.  This means that at the very
beginning, the dynamics of FOTOCs are irrelevant to the system
dynamics and affected most by the environment ($\omega_c$), and we may
call this part the pure scrambling process. However, soon after the
information of the polarization dynamics is adulterated into the
FOTOCs which makes them deviate from each other. This deviation
happens before the saturation, therefore it is not so appropriate to
identify the scrambling time as the location of the first maximum.

The pure scrambling time is hard to be identified since the
information of system dynamics is quickly adulterated. This is the
reason why we just treat the scrambling time as a vague concept in the
previous section. In the beginning, the growth of FOTOCs is fast such
that it can be treated as exponential growth. Suppose in pure
scrambling process we have
$[1-\mathcal{F}(t)]/\xi^2\sim e^{\lambda_Qt}$, then the coincidence of
$[1-\mathcal{F}(t)]/\alpha^2\xi^2$ indicates that $e^{\lambda_Qt}$ is
proportional to $\alpha^2$, and from which we can deduce that
$\lambda_Q\sim\ln\alpha$.

\section{Conclusions}
\label{sec:conclusions}

Using numerically exact time-evolving matrix product operators
approach, we study the FOTOC in the unbiased spin-boson model. It is
reported that in the Ising chain model, the OTOC will revive and
recover unity in the integrable case and oscillates in the
nonintegrable case \cite{li2017-measuring}. A similar recurrence
phenomenon is also reported in quantum Rabi model
\cite{kirkova2022-out} where the FOTOC oscillates with a certain
amplitude. Unlike the quantum Rabi model, the bath in the spin-boson
model consists of oscillators of continuous spectrum and thus the
consideration of recurrence phenomena is excluded. All FOTOCs in the
spin-boson model feature exponential growth initially and eventually
remain at some nonzero values. This indicates that the information
does scramble in the spin-boson model. Despite a factor $\alpha^{-2}$
the FOTOCs with same $\omega_c$ coincide within a very short
time. This indicates that the quantum Lyapunov exponent
$\lambda_Q\sim \ln\alpha$.

After the initial exponential growth process, the information of the
system dynamics starts to adulterate into the FOTOCs, which makes them
start to deviate from each other before the saturation. After the
scrambling time, the FOTOC shows a compressed preview of the system
dynamics in the sense that the FOTOC shows similar behavior to that of
the system dynamics in a short time.  Soon after the progress of FOTOC
is ahead of that of the system dynamics, and thus the FOTOC can be
used as a prediction of the system dynamics. Therefore we may say that
the process of the FOTOCs consists of two subprocesses that one is the
information scrambling due to the entanglement of the system and the
bath, and another is the adulteration of the information of system
dynamics. This two subprocesses compete with each other. The
scrambling dominates within a very short time and then it decays to
the saturation, and finally the information of system dynamics takes
the domination.

\end{document}